\documentclass[12pt]{article}
\usepackage{geometry}
\usepackage[utf8]{inputenc}

\usepackage{amssymb,amsmath,epsfig}
\begin{document}
\title{\textbf{Extended phase space thermodynamics of black hole with non-linear electrodynamic field}}
\author{G. Abbas$^1$ \thanks{ghulamabbas@iub.edu.pk}
and R.H. Ali$^1$ \thanks{hasnainali408@gmail.com}
\\$^1$Department of mathematics, The Islamia University of Bahawalpur,\\
Bahawalpur-63100, Pakistan.}
\date{}
\maketitle
\begin{abstract}
This paper deals with the thermodynamical properties of black hole formulated in Einstein theory of relativity and associated with a nonlinear electromagnetic field. The transition of the black hole is analyzed using the parameters mass, electric charge, coupling constant, and cosmological constant. We find-out the thermodynamical aspects of exact black hole solutions to compute the black hole mass, temperature, entropy, Gibbs free energy, specific heat and the critical exponents in the phase space. Further, we study the stability of the black hole solution by specific heat and Gibbs free energy. We look at the first and second phase changes and show a P-V criticality, which is like the Van der Waals phase change. We also examine the equation of the state and the critical exponents.
\end{abstract}
\section{Introduction}
In modern physics, black holes (BH) are considered important objects as they represent a complex system of temperature, entropy, specific heat, and enthalpy \cite{1}. These are the types of extended space that has aroused the interest of scientists for decades. The broad connection between thermodynamics, Einstein gravity, and quantum physics is evident in the formation of all four laws of BH thermodynamics. A common connection, as Einstein described, is the only time theory of the space of gravity. Within the dark holes, there is an event horizon at that point in the same direction, as no one can return or escape its gravitational pull, even at the speed of light.
To solve the problem of singularity by taking the usual model without singularity, as this model provides the appropriate division into the final phase of the fall of gravity to remove future singularity, was proposed by Sakharov \cite{2} and Gliner \cite{3}. Using this concept, Bardeen provides a solution for the BH that reveals the horizon but lacks singularity.

Various modified theories of General Relativity (GR), such as Einstein-Gauss-Bonnet theory (EGB) \cite{5}-\cite{8}, $f(T)$ gravity theory \cite{9}, $f(R)$ gravity \cite{10} and Newmann-Jeans algorithm for rotating BH \cite{11,12}, were used to obtain standard solutions of the BH. Phase modification has been used to discuss the thermodynamics of BH for a long time. The term "phase change" refers to a change in the basic state of an entity. Hawking and Page \cite{18} identified the thermodynamic features of the Anti-de-Sitter (AdS) BH in detecting the presence of phase change in the phase space. The first phase change was exposed by the BH of the charged RN-AdS\cite{19,20}.
In \cite{21}-\cite{23}, the cosmological constant successfully refers to thermodynamic pressure. This is a major source of inspiration for calculating the volume of the BH and thermodynamic pressure, as well as the re-introduction of the P-V term in BH thermodynamics \cite{24}-\cite{26}.

 The universal effects of non-linear electrodynamics (NLED) theory completely reviewed. The subject of universal evolution, as evidenced by the Born-Infeld theory \cite{13}-\cite{15} has been studied. The non-linear interaction of the electromagnetic field (EM) creates the cosmic conditions, which are necessary for the event to take place. Recent research has shown the importance of NLED in two of the most important aspects of cosmology, related to the transition time of large and very small regions. A powerful magnetic field has already been shown to win the chance for a single region in Maxwell's field on NLED. Cosmological models containing NLED have attracted a lot of attention over the last few years \cite{27}-\cite{29}. The interest in studying the NLED in the context of cosmology has grown significantly as a result of these important findings \cite{30}-\cite{40}. It was discovered that the original singularity of the Big Bang could not have been avoided if the original Universe had been ruled by non-linear magnetic field \cite{41}-\cite{49}.

  In the cosmos, NLED fields may play an important role. However, when seen from the BH, Einstein's solutions for the gravitational force and the NLED field are surprising. Because in order to comprehend such solutions, the functional relationship between nonlinear events in strong magnetic fields and electromagnetic fields must be understood. Other solutions for charged BHs and black string are present in \cite{50}-\cite{54}. It has been found that with the use of NLED field, not only the singularity of the Big Bang but also the singularity of the BH can be prevented. As a result, some common BH solutions have been found \cite{60}-\cite{65} that are regular and singularity free solutions. The Kerr-Newman-de Sitter space-time has an inner, an event, and a cosmic horizon. The BHs with many horizons in NLED fields have been investigated in Refs.\cite{66}-\cite{70}. Gunasekaran et.al \cite{72} have been studied of charged BH thermodynamics including the effect of NLED and analyze the critical behavior of charged and rotating BHs.

 This research specifically aims to investigate the thermodynamic behavior of NLED BH. The effects of coupling constant $\alpha$ have been used to discuss stability and phase transition of BH. The thermal stability has been discussed through two conditions: One isothermal pressure, and secondly heat capacity $Cv$. For this paper, the layout will look like this. In the section \textbf{II}, we present the precise review of BH solution in Einstein's gravitational field with NLED field. Also, we have discussed the thermodynamical quantities of NLED BH in the section \textbf{III}. The section \textbf{IV} deals with the $P-V$ criticals and critical parameters. The last portion, which summaries the results of our findings.
\section{Review of Nonlinear Electrodynamic BH Solution}
In this section, we will examine the NLED BH solution \cite{71} that comes from the action of NLED theories that have been coupled with gravity
\begin{equation}
S= \frac{1}{16 \pi} \int \sqrt{-g}\Big(R + K(\psi)\Big) d^{4}x,
\end{equation}
\\ where
\begin{equation}\label{F}
\psi = F_{\mu\nu} F^{\mu\nu}, \,\,\,  F_{\mu\nu}=  A_{\nu;\mu}-  A_{\mu;\nu}.
\end{equation}
Here $A_{\mu}$ gives the Maxwell field, Ricci scalar $R$ and $\psi$ function $K(\psi)$.
The field equations of Einstein given by the change of action relating to the metric,
 \begin{equation}
 G_{\mu \nu} = -2K_{, \psi} F_{\mu \lambda} F_{\nu}^{\lambda}+ \frac{1}{2}g_{\mu\nu}K,\quad\quad K_{, \psi}\equiv \frac{dK}{d\psi}.
\end{equation}
The Maxwell general equations are determined by the field,
\begin{equation}
(K_{, \psi}
 F^{\mu\nu})_{;\mu}=0.
\end{equation}
The spherically symmetric and static background metric. \cite{71} is given by,
\begin{equation}
ds^{2}= -N(r)dt^{2}+ \frac{1}{N(r)}dr^{2}+ f(r)^{2}d\Omega_{2}^{2},
\end{equation}
The only non-zero component of the Maxwell field tensor $A_{\mu}$  is given by
\begin{equation}
A_{0}= -\phi(r).
\end{equation}
\begin{equation}
\psi = -2\varphi'^{2},
\end{equation}

The field equations by Einstein and the generalized equations by Maxwell \cite{71} were also handled,
\begin{equation}
\frac{-N'f'}{f}-\frac{2N f''}{f} + \frac{1}{f^{2}}-\frac{N f'^{2}}{f^{2}}= 2K_{,\psi} \phi'^{2}+ \frac{1 }{2}K,\label{a1}
\end{equation}
\begin{equation}
\frac{-N'f'}{f}+ \frac{1}{f^{2}}-\frac{N f'^{2}}{f^{2}} = 2K_{,\psi} \phi'^{2}+ \frac{1 }{2}K,\label{a2}
\end{equation}
\begin{equation}
\frac{-N'f'}{f}+\frac{Nf''}{f} +\frac{1}{2} N'' = - \frac{1 }{2}K,\label{a3}
\end{equation}
\begin{equation}
(f''K_{, \psi}\phi')' = 0.\label{a4}
\end{equation}
The differentiation with respect to $r$ is represented by the prime.
The Maxwell field equation (\ref{a4}) defines electric charge,
\begin{equation}
Q = f^{2} \phi' K_{, \psi}.
 \end{equation}
 From the difference of Eqs.(\ref{a1}) and Eqs. (\ref{a2}), we get
 \begin{equation}
f''(r)=0.
 \end{equation}
 As a result, the solution to $f$ is
 \begin{equation}
f(r)=r.
 \end{equation}
 We assume the Maxwell Lagrangian $K$ form to solve the field equations,
\begin{eqnarray}
K= 2\sqrt{2\alpha}\sqrt{-\psi}-\psi- 2\Lambda,
\end{eqnarray}
where $\alpha$ has the dimension of length. The theory reduces to Maxwell's When $\alpha=0$, (see for detail Eq. (\ref{F}) and $ 2\Lambda $ is  cosmological constant factor.
Ultimately,
\begin{eqnarray}
\phi=r\sqrt{\alpha}+\frac{Q}{r},
\end{eqnarray}
\begin{equation}
N(r)= 1-\frac{2 M}{r}+\frac{Q^2}{r^2} +\frac{r^2}{l^2}+2 \sqrt{\alpha } Q-\frac{\alpha  r^2}{3}. \label{l1}
\end{equation}
  Here $Q$, $\alpha$ and $\Lambda$ are electric charge, coupling constant and cosmological constant, respectively. If one vanishing $\alpha\rightarrow 0$, its nothing but just RN-AdS solution and $Q=0$ reveals the Schwarzschild AdS BH solution.
\section{Thermodynamics Quantities of BH}
Now we are going to discuss about the thermodynamical properties of NLED BH \cite{71}.
\begin{figure}[ht!]
\centering
\begin{minipage}[b]{0.4\textwidth}
\includegraphics[width=3.1in, height=3in]{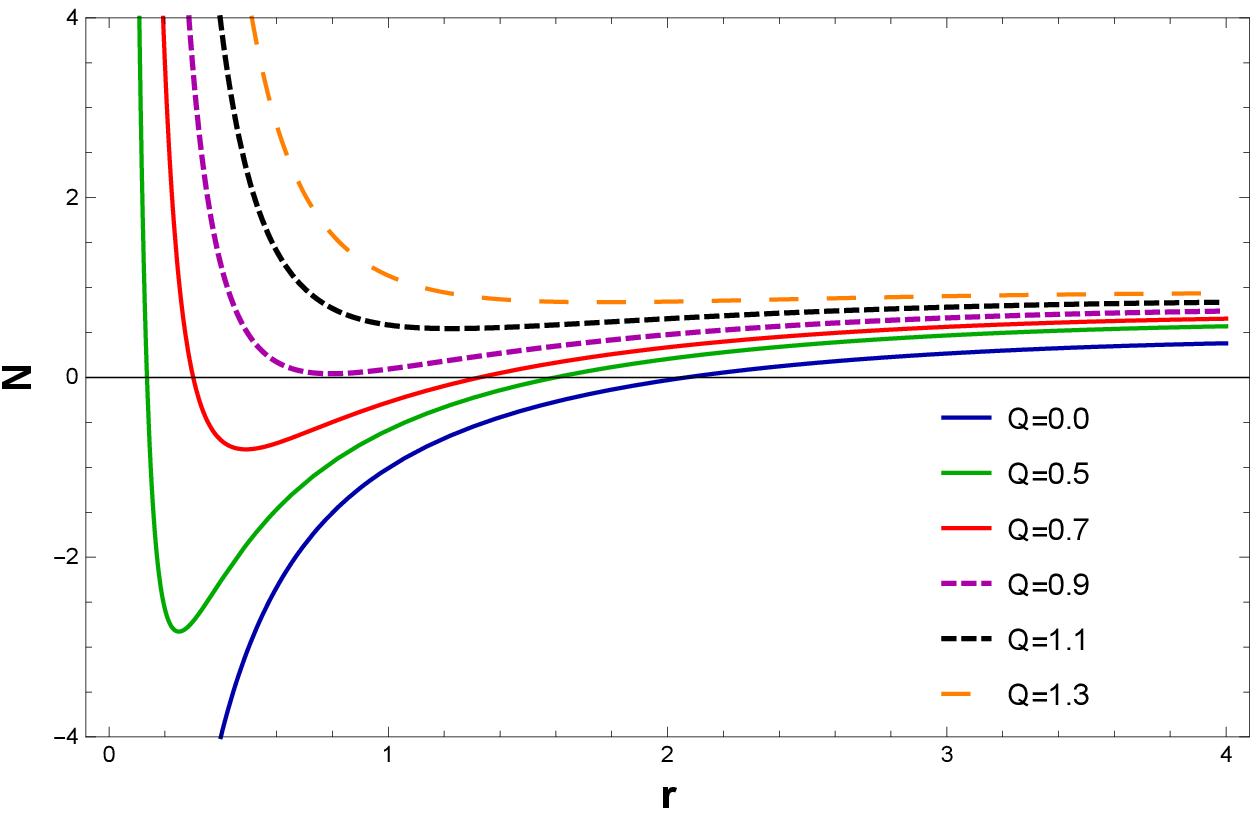}\
\caption{The graph of $N$ as a function of $r$ for $\alpha=0.030$,  $l=20$ and $M=1$}
\end{minipage}
\hfill
\begin{minipage}[b]{0.4\textwidth}
\includegraphics[width=3.1in, height=3in]{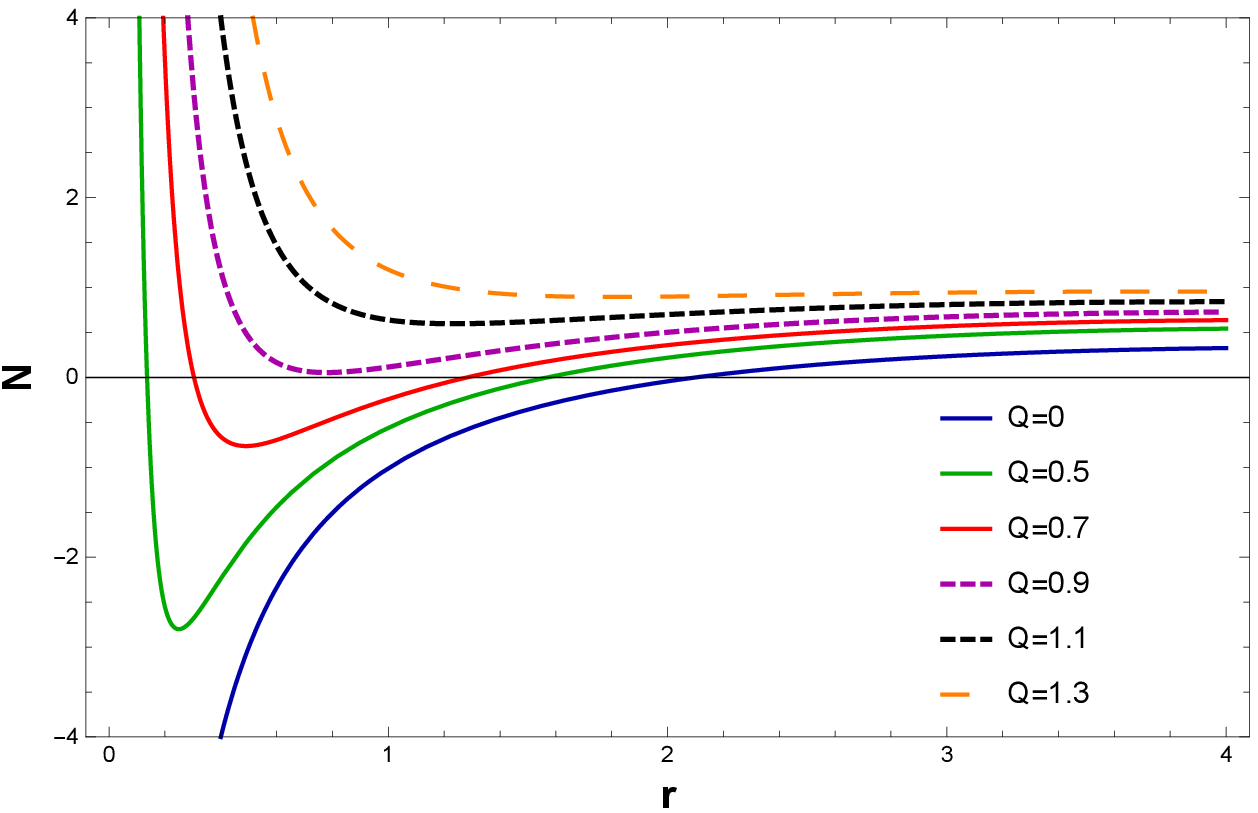}
\caption{The graph of $N$ as a function of $r$ for $\alpha=0.040$, $l=20$ and $M=1$}
\end{minipage}
\begin{minipage}[b]{0.4\textwidth}
\includegraphics[width=3.1in, height=3in]{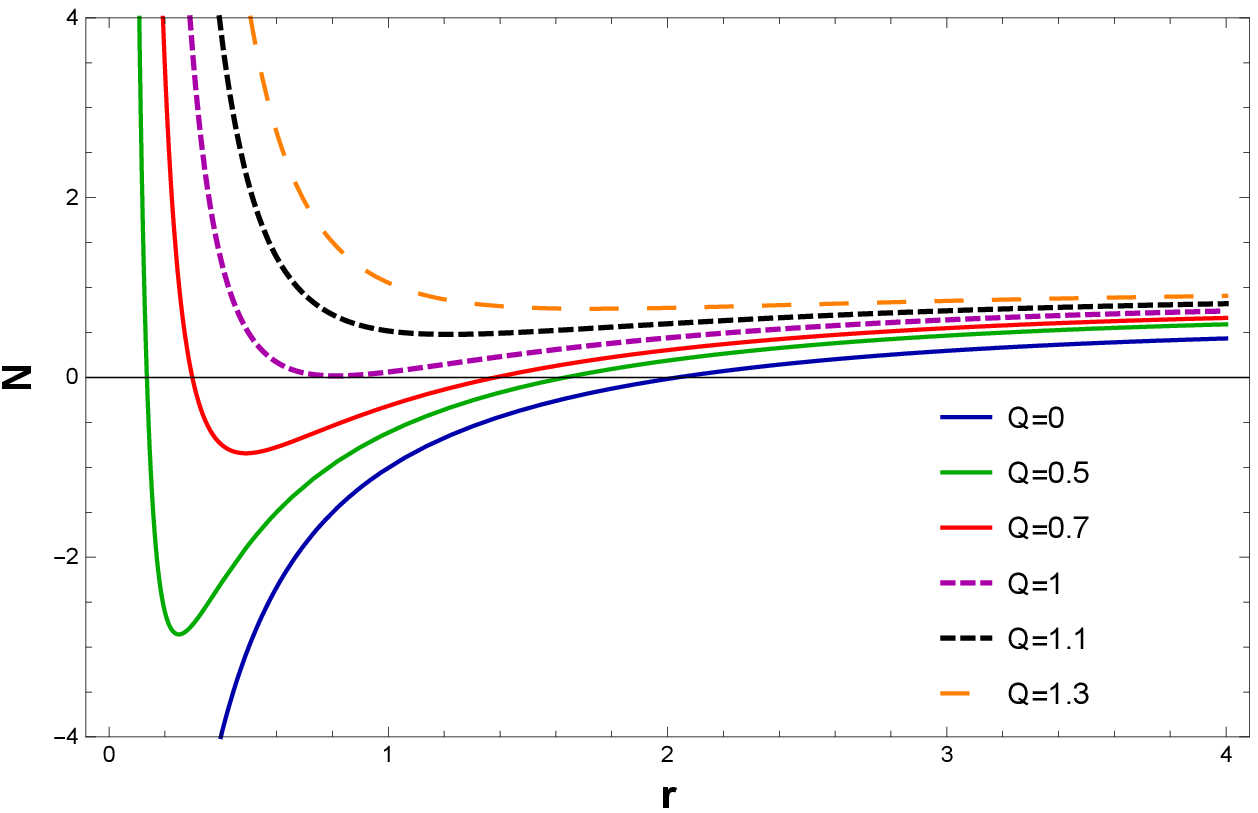}
\caption{The graph of $N$ as a function of $r$ for $\alpha=0$, $l=20$ and $M=1$}
\end{minipage}
\end{figure}

 Using  Eq. (\ref{l1}), that $N(r)=0$, in  one may determine the physical characteristics of a BH event horizon. The equations are solved by changing the values of the parameters $Q$, $\alpha$, and $l=20$. The behavior of the discovered solution for a fixed value of $\alpha$ is depicted in the figures $1$, $2$ and $3$. It is evident that as the charge $Q$ is increased, the BH size grows.
From the Figs. \textbf{1} and \textbf{2}, we can deduce from the answer that it shows a naked singularity at $Q=1.1$ and $ Q=1.2$, which means that it gives no event horizon, hence no black holes, when $Q>Q_{c}$, one double zero, which corresponds to an extreme BH, for $Q=Q_{c}$, and two simple zeros, corresponding to two event horizons, which denote a non-extreme BH for $Q<Q_{c}$. In plot 3, as $\alpha=0$ and $Q=0$ demonstrate event horizons of RN-BH as well as Schwarzschild BH respectively. There are two categories of BH solutions that appear, associated with extremal and non-extremal.

Taking $N(r_+)=0$ enables us to determine BH mass as follows
\begin{equation}
M=\frac{1}{2} r_{+} \left(\frac{r_{+}^2}{l^2}+\frac{Q^2}{r_{+}^2}+2 \sqrt{\alpha } Q-\frac{\alpha  r_{+}^2}{3}+1\right).\label{a5}
\end{equation}
We get the BH temperature from $T=\frac{k}{2\pi}=\frac{N'(r_{+})}{4\pi}$,
where $k$ is the surface gravity.
Thus temperature takes the following form
\begin{equation}
T=\frac{1}{{4 \pi r_{+}}}\Big(1+\frac{3 r_{+}^2}{l^2}-\frac{Q^2}{r_{+}^2}+2 \alpha  Q-\alpha  r_{+}^2\Big).\label{a6}
\end{equation}
\begin{figure}[ht!]
\centering
\begin{minipage}[b]{0.4\textwidth}
\includegraphics[width=3.1in, height=3in]{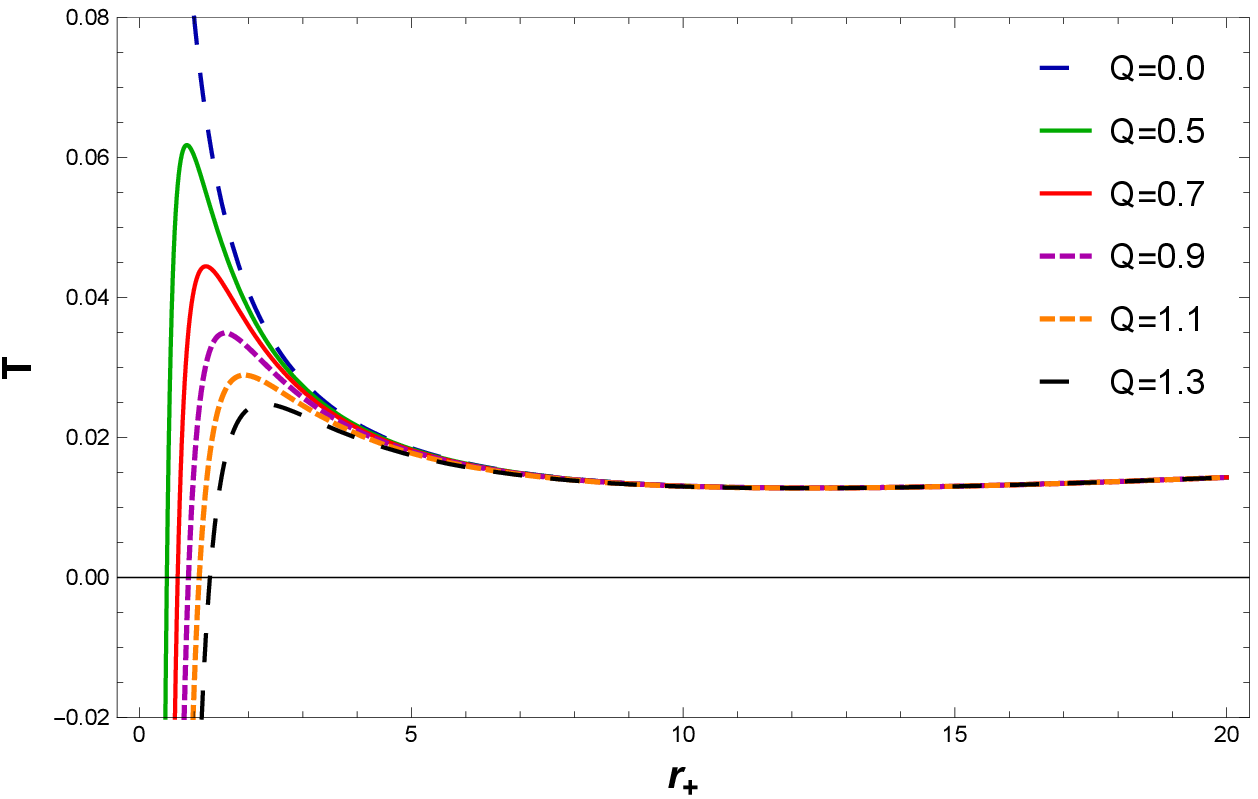}
\caption{The graph of $T$ associated with $r_{+}$ for $\alpha=0.0020$, and $l=20$}
\end{minipage}
\hfill
\begin{minipage}[b]{0.4\textwidth}
\includegraphics[width=3.1in, height=3in]{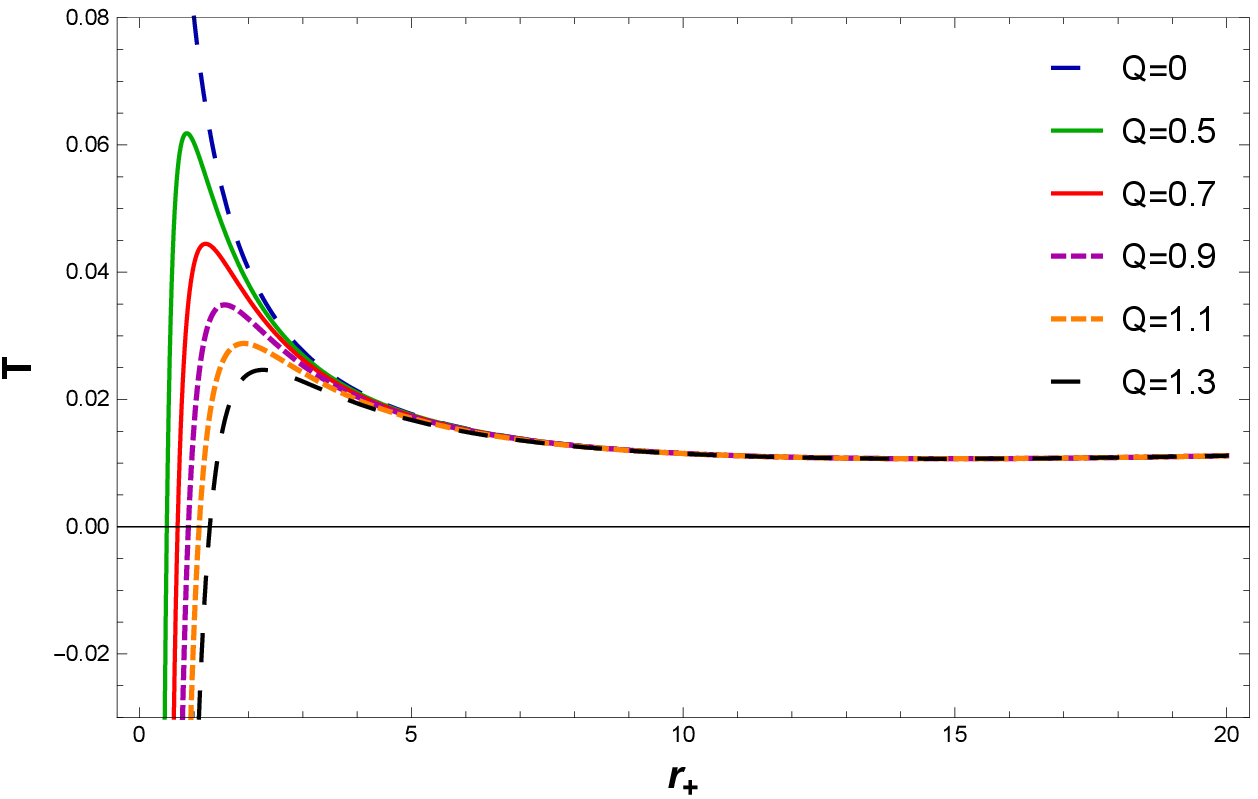}
\caption{The graph of $T$ associated with $r_{+}$ for $\alpha=0.0030$, and $l=20$}
\end{minipage}
\hfill
\begin{minipage}[b]{0.4\textwidth}
\includegraphics[width=3.1in, height=3in]{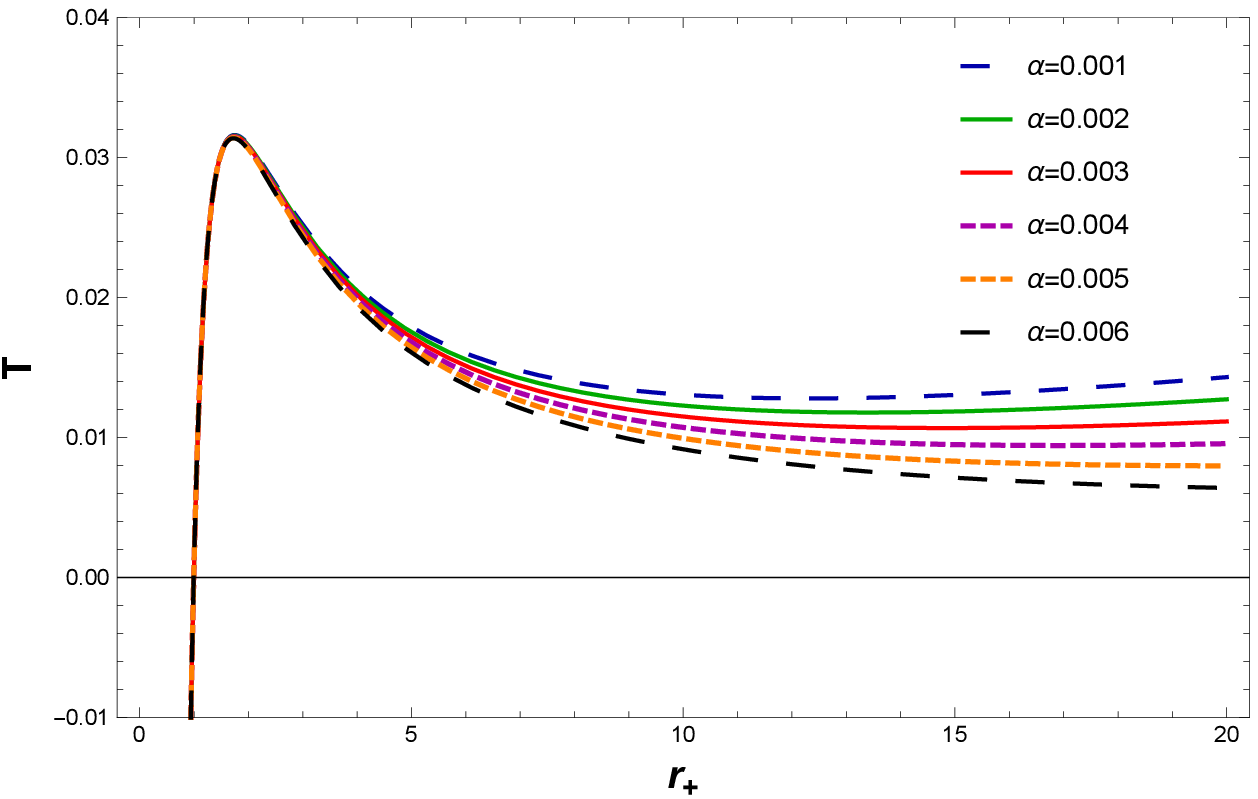}
\caption{The graph of $T$ against $r_{+}$ for $Q=1$, and $l=20$}
\end{minipage}
\hfill
\begin{minipage}[b]{0.4\textwidth}
\includegraphics[width=3.1in, height=3in]{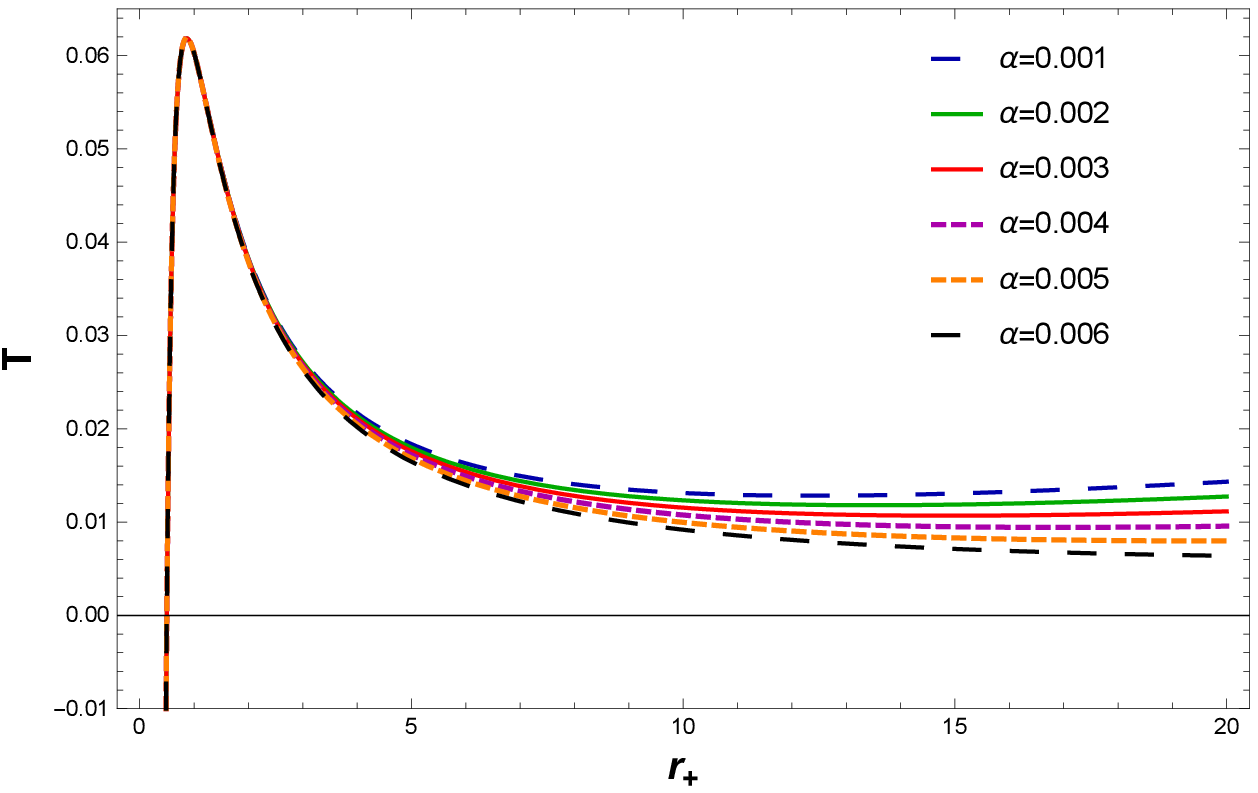}
\caption{The graph of $T$ against $r_{+}$ for $Q=0.5$, and $l=20$}
\end{minipage}
\end{figure}

In Figs.\textbf{4}, \textbf{5}, \textbf{6}, and \textbf{7}, we plot the NLED charged BH Hawking temperature, for the small radius BH, as well as large radius BH, for a variety of charge values, $Q$ and the coupling constant $ \alpha$ with $l=20 $. From figures, \textbf{4}, and \textbf{5}, it is clear that Hawking temperature of BH increases (decreases) to maximum (minimum) value when charge $ Q $ decreases (increases) with the increase of the horizon radius. It is worthy to mention that, with fixed  NLED coupling constant $ \alpha=0.0020, 0.0030$, and electric charge start vary from $Q=0$ to $Q=1.3,$ explain the Hawking temperature of Schwarzschild and RN BHs. The Figs. \textbf{6}, and \textbf{7}, reveal the behavior of Hawking temperature with fixed values of charge $Q=0.5$, and $1$ and NLED coupling parameter vary from $\alpha=0.0010$ to $\alpha=0.0060$, and describe the increasing behavior of Hawking emperature as the coupling parameter decreases with the increase of horizon radius (from small to large BH) and exhibit the universal pattern of stability. Hence the system remain stable for small as well as large BH.

The BH entropy is derived from $dM=TdS,$ which integrates to give
$S=\int\frac{dM}{T}dr_{+},$

\begin{equation}
S=\pi r_{+}^{2}.\label{a7}
\end{equation}
The heat capacity can be defined as $ C=\frac{\partial M}{\partial T}$, which simplifies to
\begin{equation}
C=\frac{2 \pi  r_{+}^2 \left(l^2 \left(Q^2-2 \sqrt{\alpha } Q r_{+}^2+\alpha  r_{+}^4-r_{+}^2\right)-3 r_{+}^4\right)}{l^2 \left(-3 Q^2+2 \sqrt{\alpha } Q r_{+}^2+\alpha  r_{+}^4+r_{+}^2\right)-3 r_{+}^4}.
\end{equation}

\begin{figure}[ht!]
\centering
\begin{minipage}[b]{0.4\textwidth}
\includegraphics[width=3.1in, height=3in]{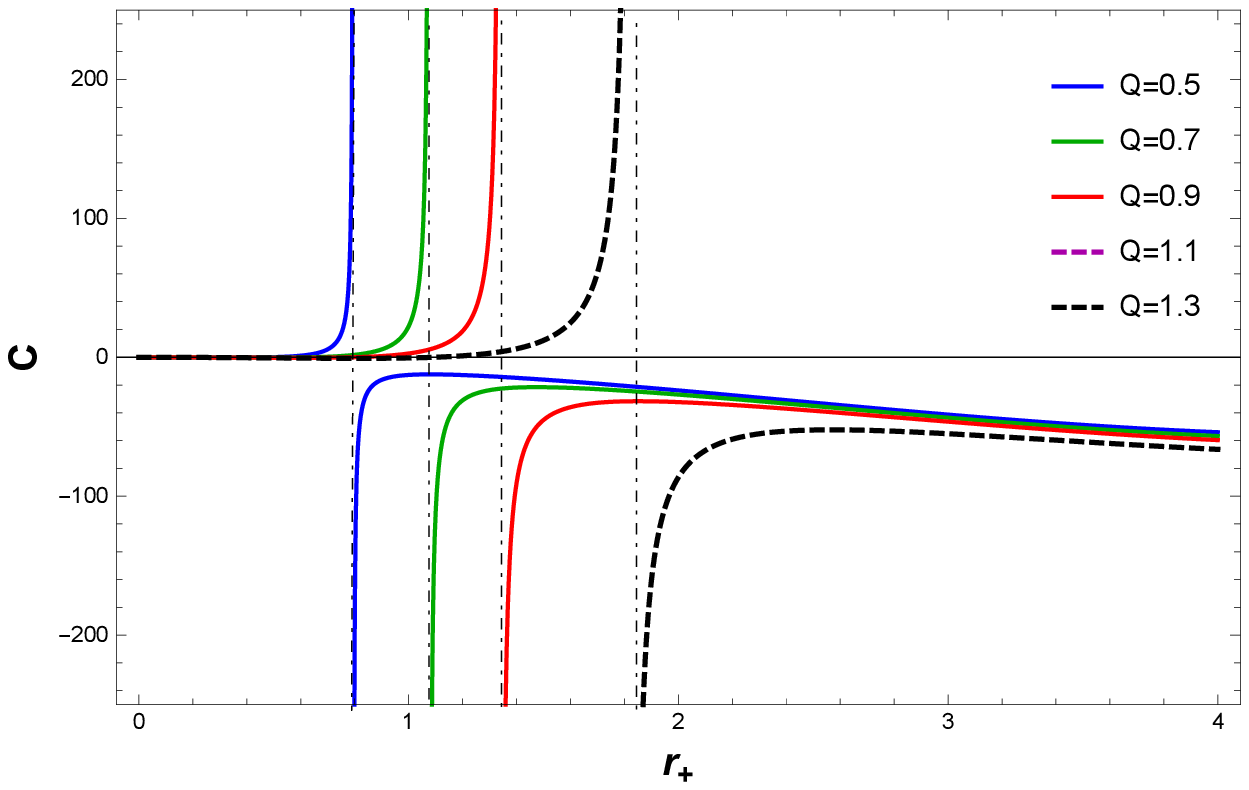}
\caption{The graph of $C$ associated with $r_{+}$ for $\alpha=0.030$ and $l=20$}
\end{minipage}
\hfill
\begin{minipage}[b]{0.4\textwidth}
\includegraphics[width=3.1in, height=3in]{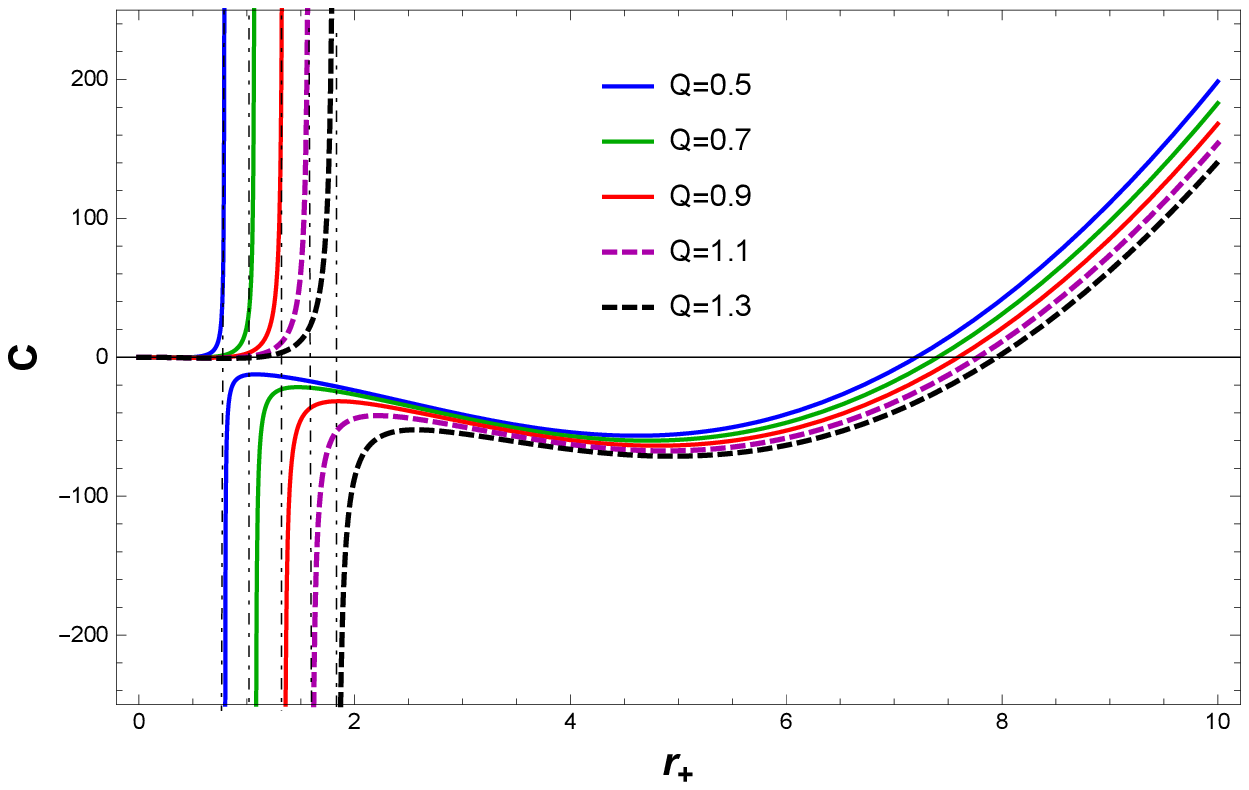}
\caption{The graph of $C$ associated with $r_{+}$ for $\alpha=0.030$ and $l=20$ }
\end{minipage}
\hfill
\begin{minipage}[b]{0.4\textwidth}
\includegraphics[width=3.1in, height=3in]{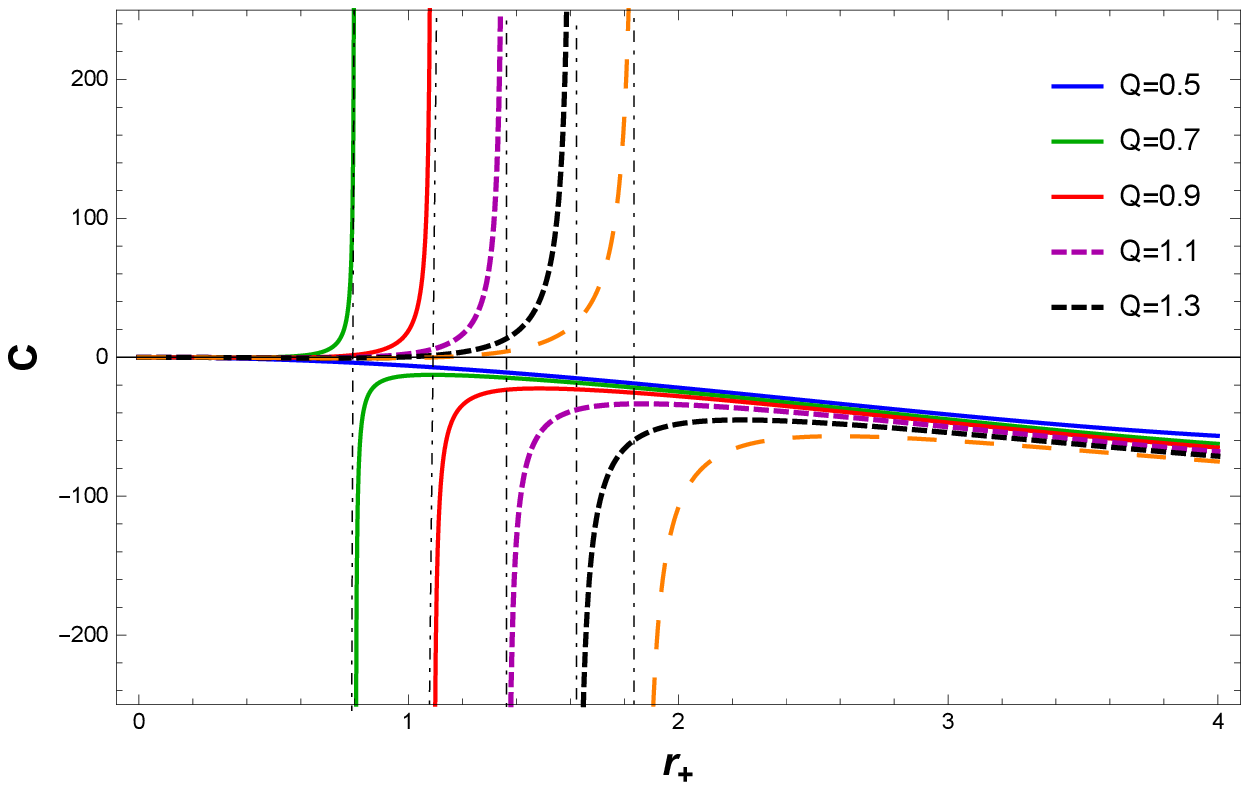}
\caption{The graph of $C$ associated with $r_{+}$ for $Q=1$ and $l=20$ }
\end{minipage}
\hfill
\begin{minipage}[b]{0.4\textwidth}
\includegraphics[width=3.1in, height=3in]{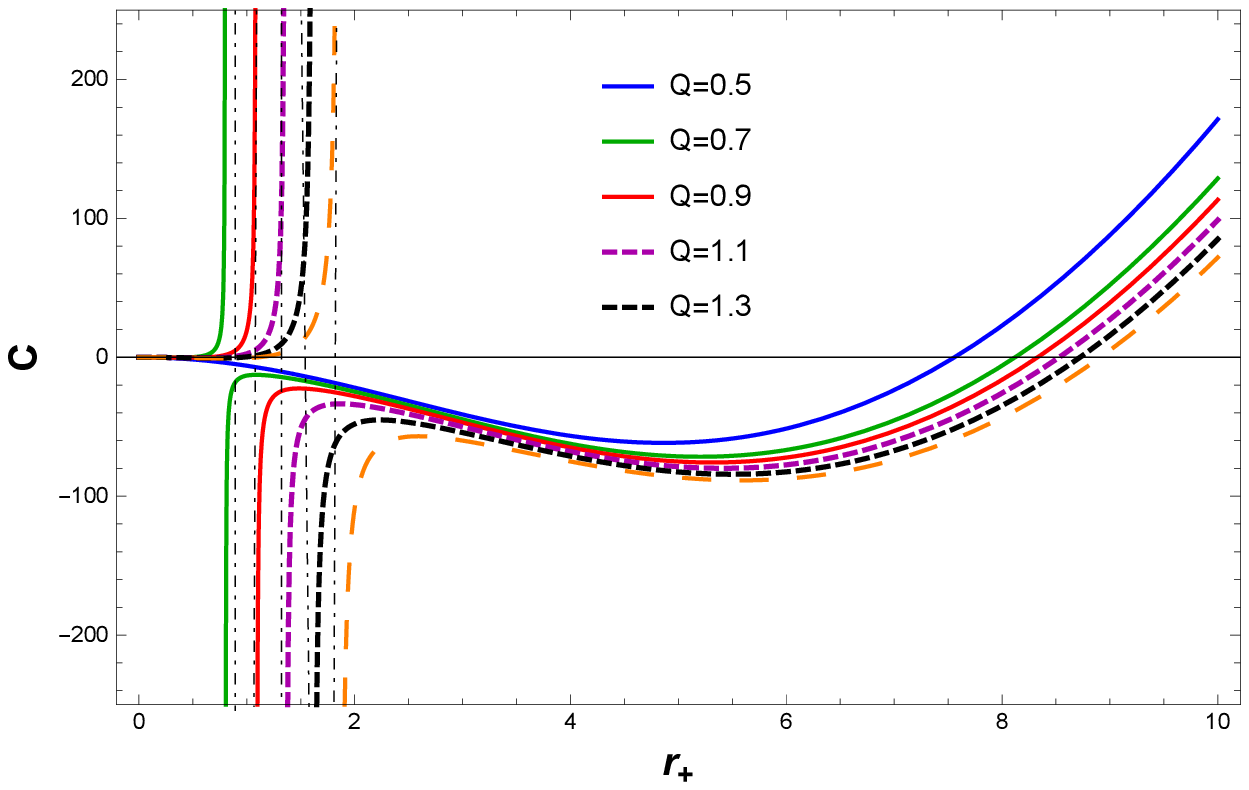}
\caption{The graph of $C$ associated with $r_{+}$ for $Q=1$ and $l=20$ }
\end{minipage}
\end{figure}
\begin{figure}[ht!]
\centering
\begin{minipage}[b]{0.4\textwidth}
\includegraphics[width=3.1in, height=3in]{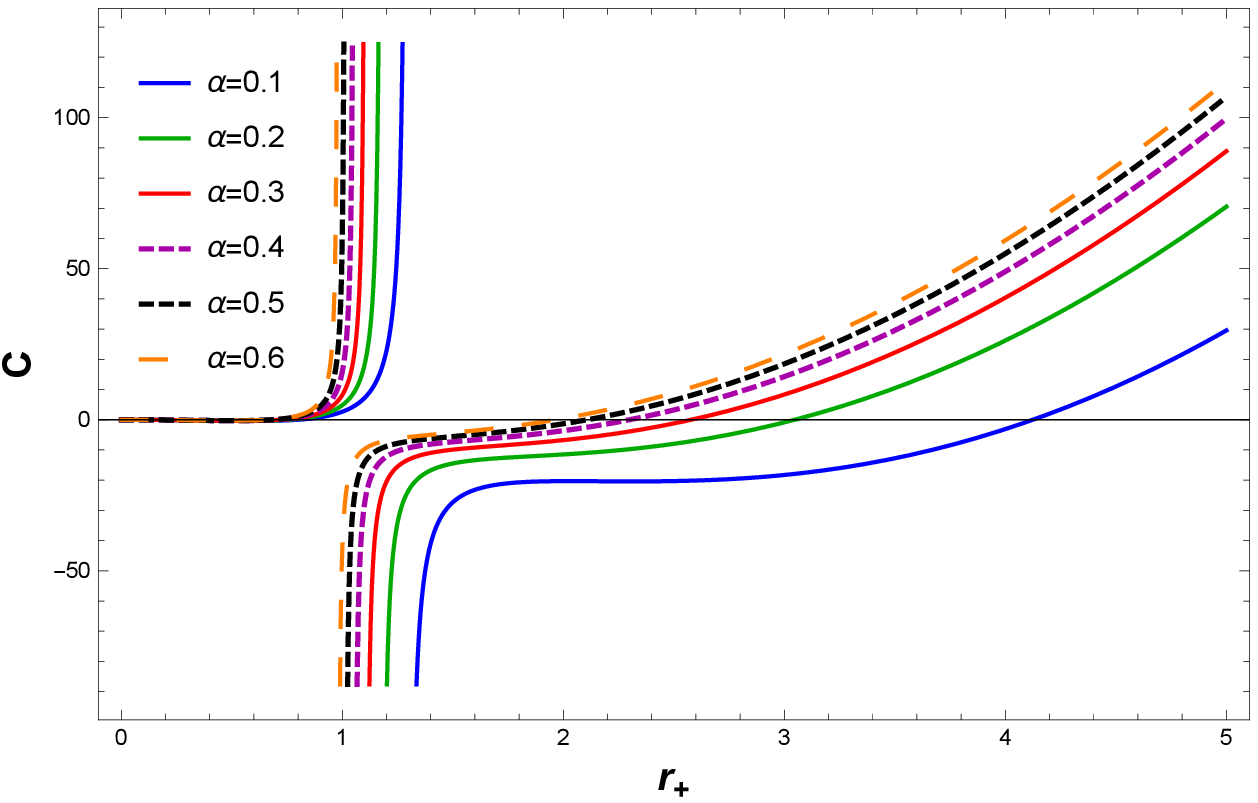}
\caption{The graph of $C$ associated with $r_{+}$ for $Q=1$ and $l=20$ }
\end{minipage}
\hfill
\begin{minipage}[b]{0.4\textwidth}
\includegraphics[width=3.1in, height=3in]{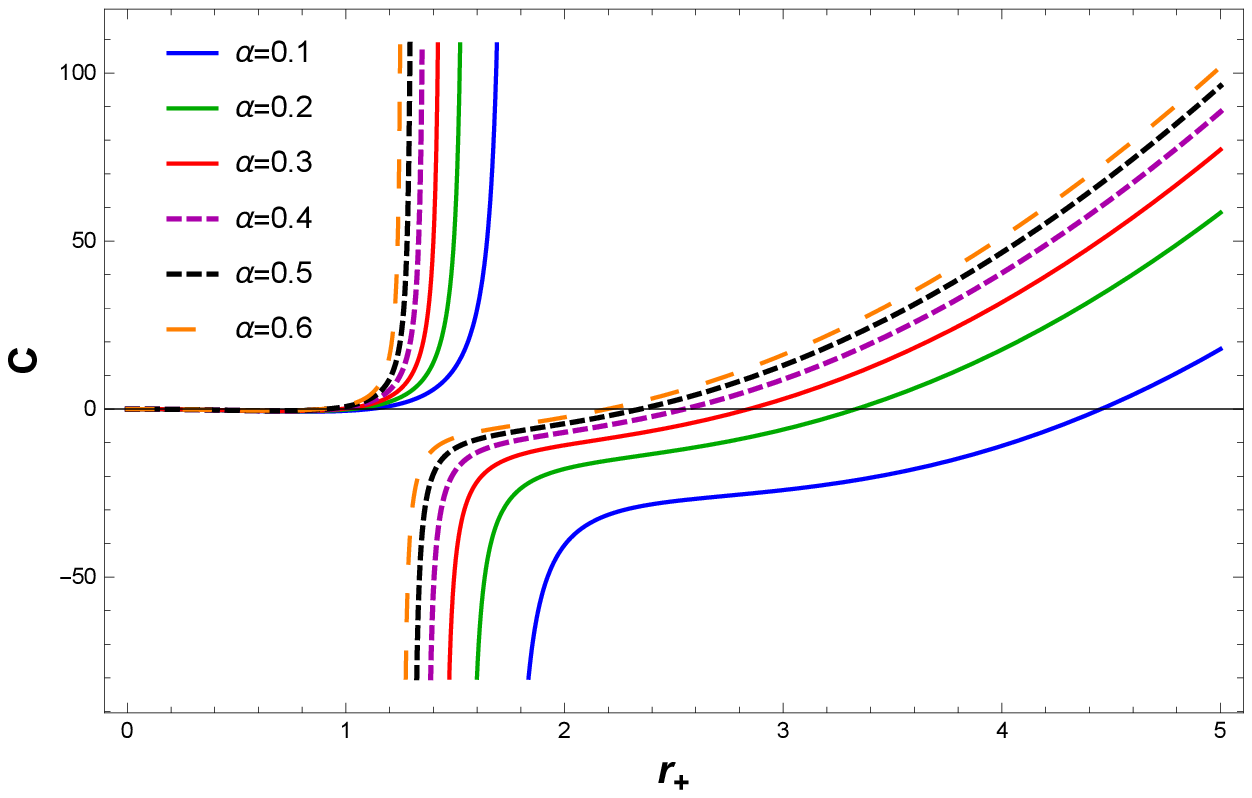}
\caption{The graph of $C$ associated with $r_{+}$ for $Q=1.5$ and $l=20$ }
\end{minipage}
\end{figure}
 We will now analyze the thermodynamical consistency of the BH solution to Einstein gravity with the NLED field. For this purpose, we turn our attention by reckoning that heat capacity indicates BH solution thermal stability. The thermodynamic stability of the BH is contingent on the behavior of the heat capacity. If $ C > 0,$ then thermodynamical system is locally stable, whereas it become unstable when specific heat $C < 0$. To identify the stability of the specific heat, the graphical representation as shown in Figs. \textbf{8}, \textbf{9}, \textbf{10} and \textbf{11}. We have plotted specific heat $C$ verses $r_{+}$ for different amounts of electric charge $Q $ and NLED coupling constant $ \alpha.$ If the electric charge $Q=0$, it gives the heat capacity of Schwarzschild AdS BH. Corresponding to the maximum temperature, critical values discontinue heat capacity curves $r_{+} = r_{c}$. At critical values, the specific heat diverges as shown in Figs.\textbf{8} and \textbf{10}. We find that as the electric charge grows, so does the critical radius. In Figs. \textbf{9}, and \textbf{11}, express the stability at initial phase and the second phase change occurs from unstable region to stable region for large BH with varying the charge $Q$ with the fixed values of NLED parameter $\alpha$. A similar behavior can be seen from Figs. \textbf{12}, and \textbf{13}, by varying NLED parameter with fixed value of charge parameter $Q$. As a consequence, the large BH in the presence of charge and coupling parameter tends to be in the stable region as compared to small BH.

Free energy is also known as Gibbs free energy(GFE), using thermodynamical quantities temperature, mass and entropy of BH, the relation $$ G= M-TS,$$ takes the form
\begin{equation}
G= \frac{1}{12} \Big(r_{+}^3(\alpha-8\pi P)+\frac{9Q^2}{r_{+}}-6\alpha-2\sqrt{\alpha}Qr_{+}+3r_{+}\Big).
\end{equation}
\begin{figure}[ht!]
\centering
\begin{minipage}[b]{0.4\textwidth}
\includegraphics[width=3.1in, height=3in]{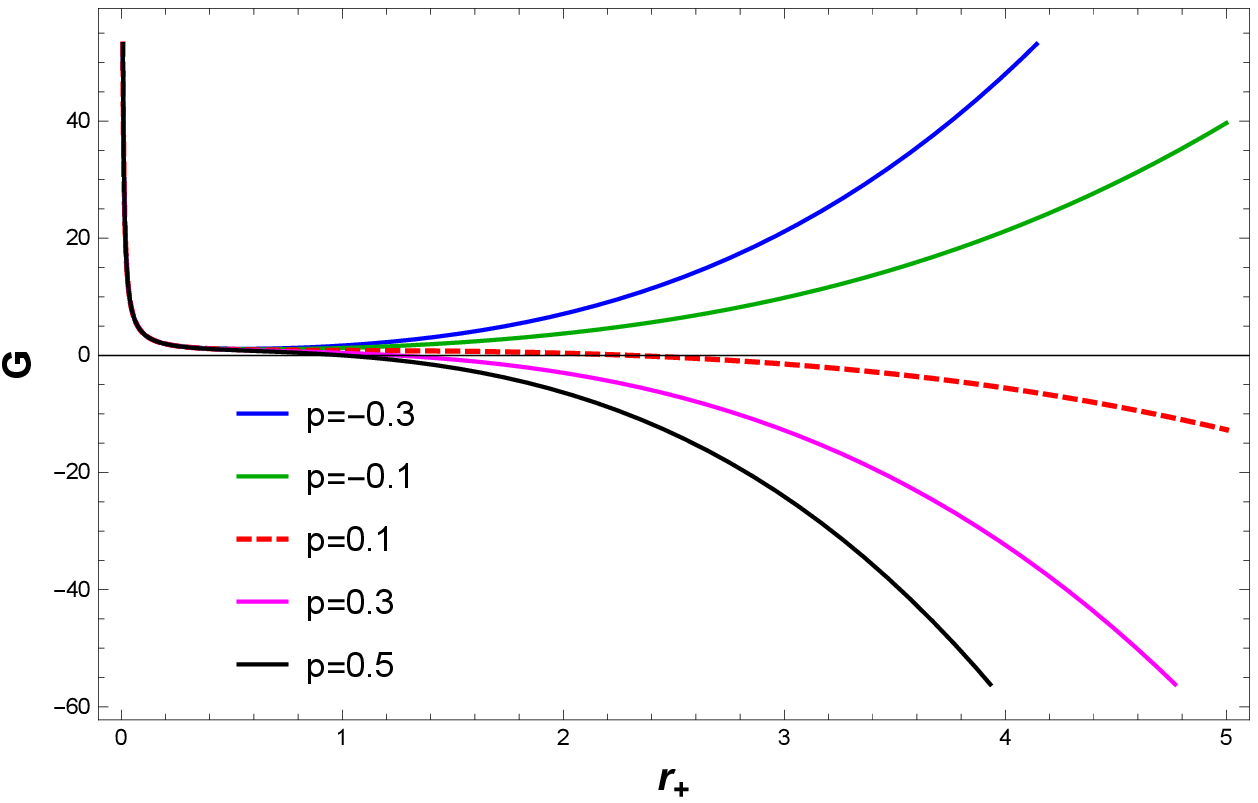}
\caption{The graph of $G$ as a function of $r_{+}$ for $\alpha=1$, $l=20$ and $Q=0.70$}
\end{minipage}
\hfill
\begin{minipage}[b]{0.4\textwidth}
\includegraphics[width=3.1in, height=3in]{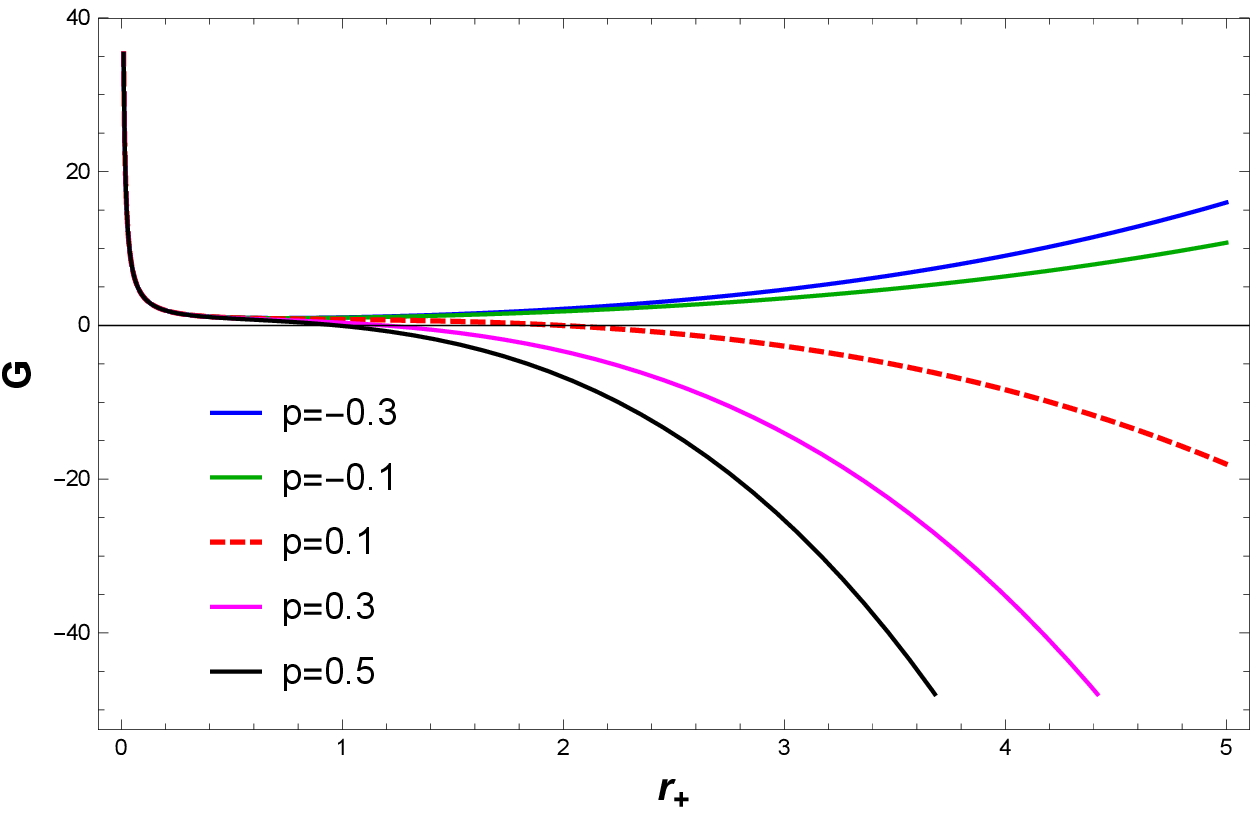}
\caption{The graph of $G$ as a function of $r_{+}$ for $\alpha=0.50$, $l=20$ and  $Q=0.70$}
\end{minipage}
\end{figure}
 To recognize the thermodynamic stability of BH, we must investigate the GFE interactions. The effect of BH parameters plays an effective part in the stability of transition. To analyze the behavior of GFE, we draw a plot of GFE $G$ as a function of BH temperature. By varying the electric charge $Q$ and fixing $ \alpha= 1 $ and $\alpha= 0.5$. It is clear from the figures \textbf{14} and \textbf{15}, that the curves associated with different values of pressure. The positive region of GFE shows stability while negative region of GFE indicate thermodynamically unstable. Hence the BHs with negative GFE release energy to the surrounding to obtain low energy state and BHs with positive GFE are globally stable.

\section{P-V Criticals and Critical Exponents}

The extensive state space pressure takes the form,
$$ P= \frac{-\Lambda}{8\pi} = \frac{3}{8\pi l^{2}},$$
using Eq. (\ref{a6}), we have
\begin{equation}
P(r_{+}, T)=  \Big(\frac{\alpha }{8 \pi }+\frac{Q^2}{8 \pi  r_{+}^4}-\frac{\alpha  Q}{4 \pi  r_{+}^2}-\frac{1}{8 \pi  r_{+}^2}+\frac{T}{2 r_{+}}\Big),\,\,\,\,\,\,  v=2r_{+}.\label{a8}
\end{equation}
Also,
\begin{equation}
P(v, T)= \Big(\frac{\alpha}{8\pi}+\frac{2Q^2}{\pi v^4}-\frac{\alpha Q}{\pi v^2}+\frac{T}{v}-\frac{1}{2\pi v^2}\Big).\label{a9}
\end{equation}
To find the critical values, we put
\begin{equation}
\frac{\partial P}{\partial r_{+}}|_{T} =0 \,\,\,\,\,\,\,\,\,\, \and \,\,\,\,\,\,\,\,\,\,  \frac{\partial^{2} P}{\partial r_{+}^{2}}|_{T} =0.
\end{equation}

\begin{figure}[ht!]
\centering
\begin{minipage}[b]{0.4\textwidth}
\includegraphics[width=3.1in, height=3in]{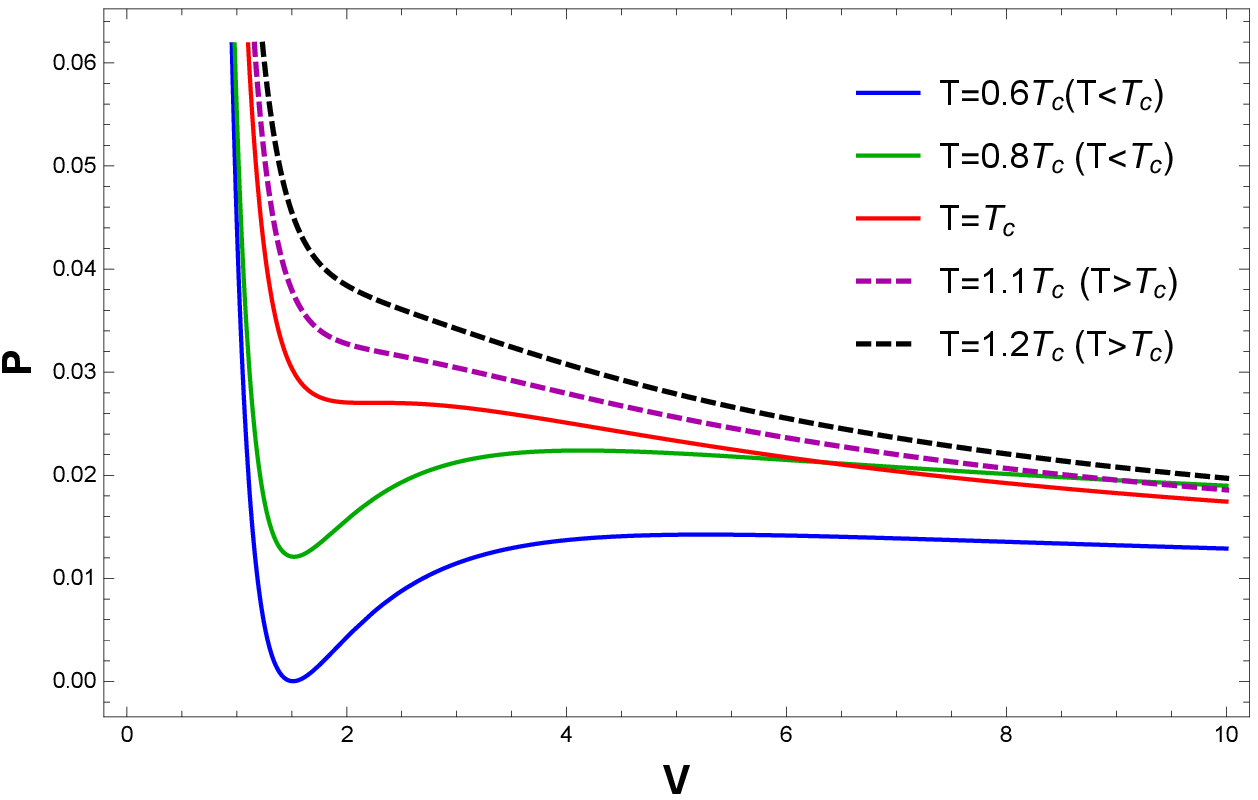}
\caption{The graph of $P$ as a function of $V$ for $Q=0.50$ and $ \alpha=0.30$}
\end{minipage}
\hfill
\begin{minipage}[b]{0.4\textwidth}
\includegraphics[width=3.1in, height=3in]{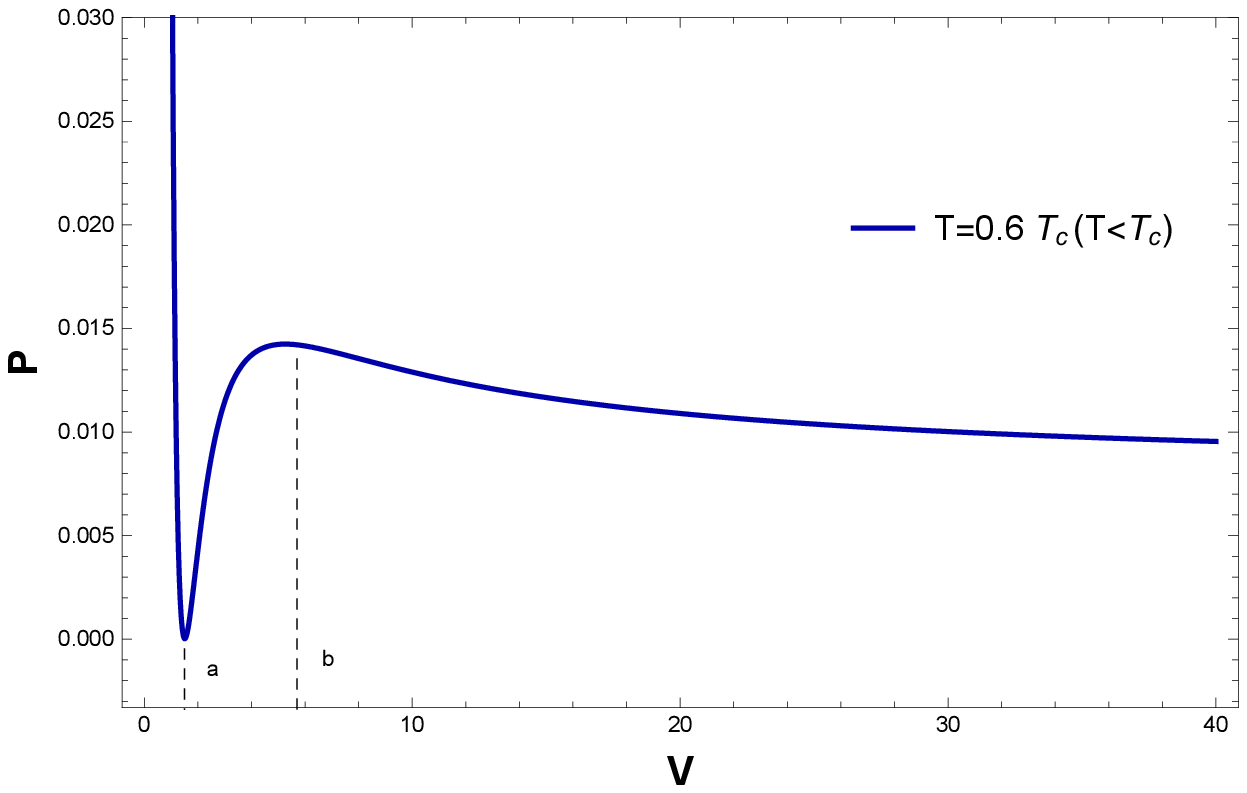}
\caption{The graph of $P$ as a function of $V$ for $Q=0.50$, $ \alpha=0.20$ and $T=0.06828$}
\end{minipage}
\hfill
\begin{minipage}[b]{0.4\textwidth}
\includegraphics[width=3.1in, height=3in]{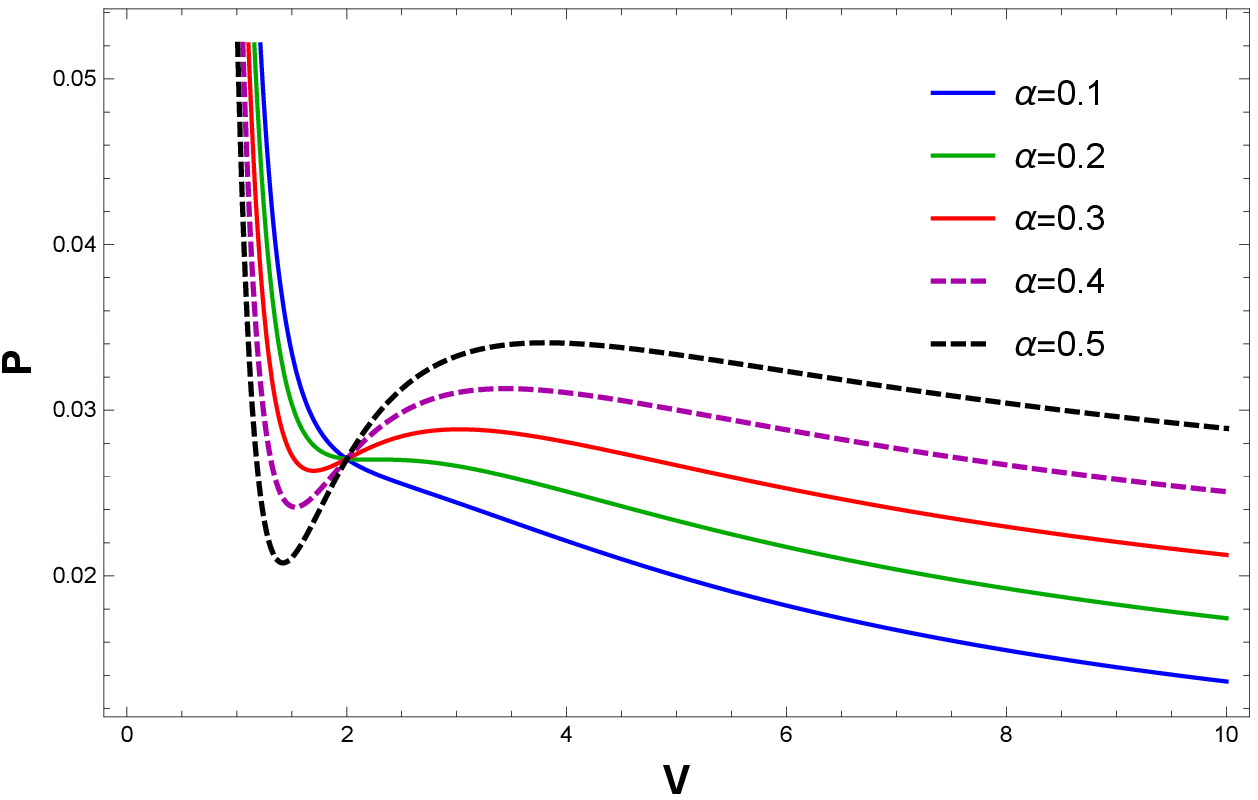}
\caption{The graph of $P$ as a function of $V$ for $Q=0.50$ and $T=0.1138$}
\end{minipage}
\end{figure}

\begin{figure}[ht!]
\centering
\begin{minipage}[b]{0.4\textwidth}
\includegraphics[width=3.1in, height=3in]{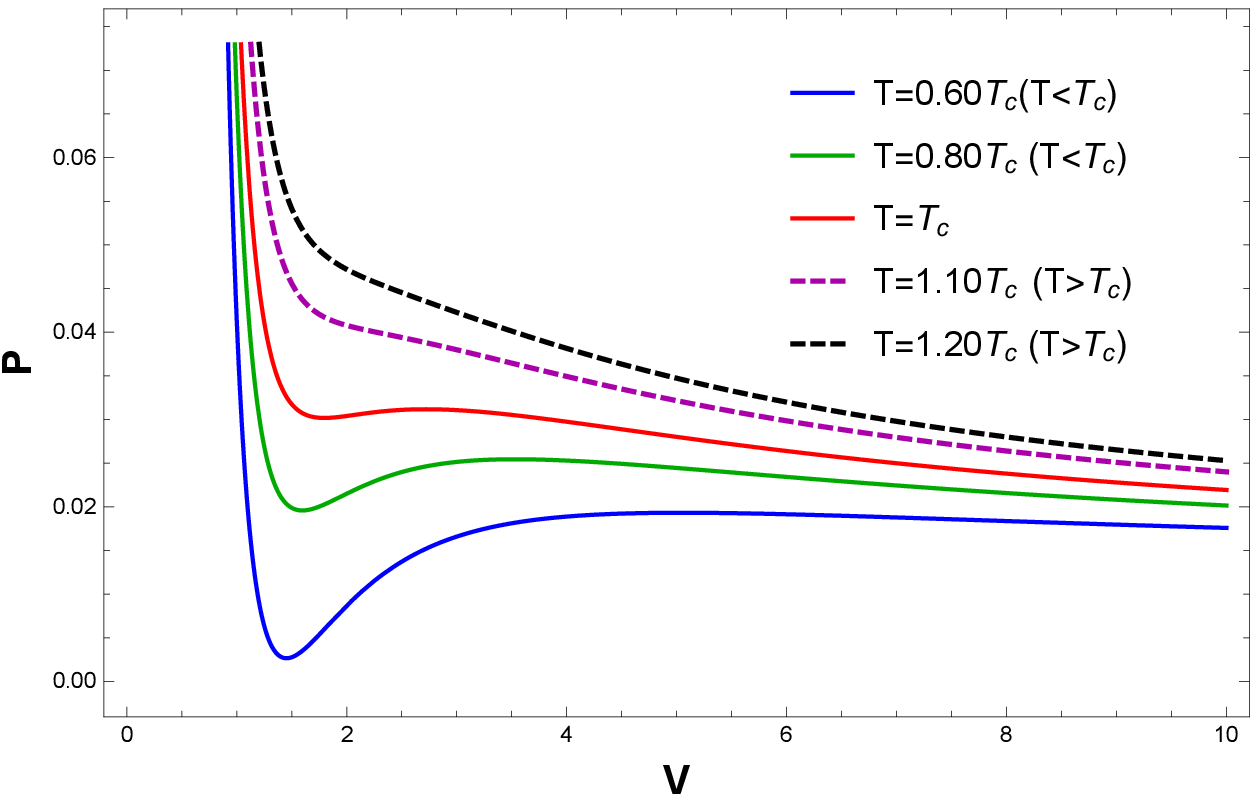}
\caption{The graph of $P$ as a function of $V$ for $Q=0.50$ and $ \alpha=0.40$}
\end{minipage}
\hfill
\begin{minipage}[b]{0.4\textwidth}
\emph{}\includegraphics[width=3.1in, height=3in]{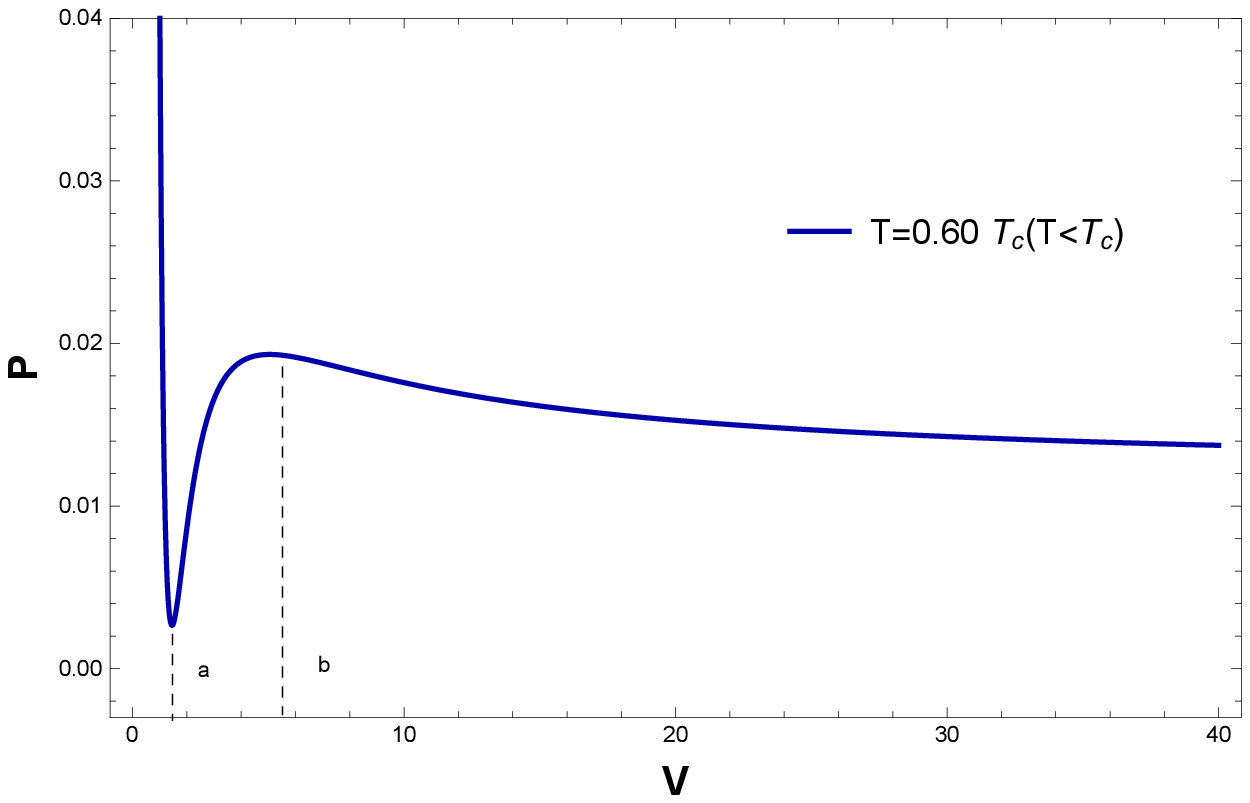}
\caption{The graph of $P$ as a function of $V$ for $Q=1$, $ \alpha=0.30$ and $T=0.07704$}
\end{minipage}
\hfill
\begin{minipage}[b]{0.4\textwidth}
\includegraphics[width=3.1in, height=3in]{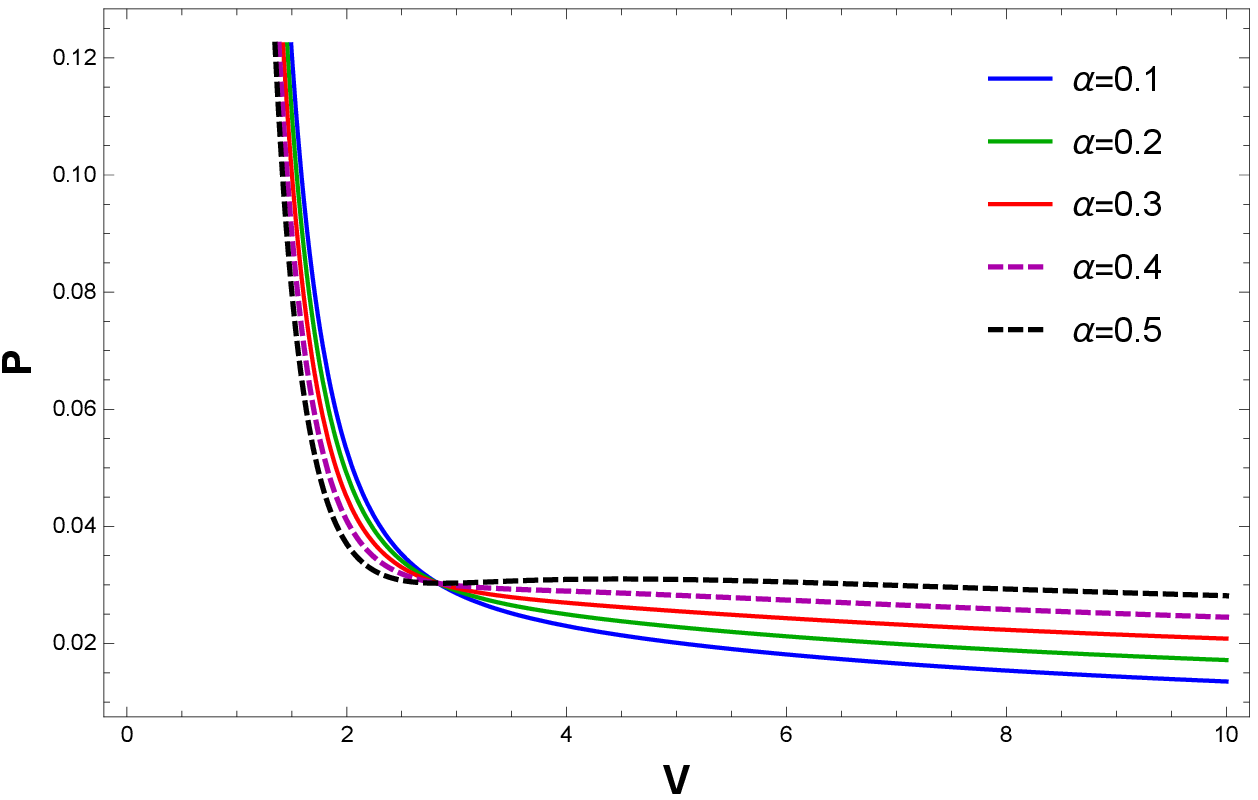}
\caption{The graph of $P$ as a function of $V$ for $Q=1$ and $T=0.1138$}
\end{minipage}
\end{figure}
 To comprehend the phase transition of the BH solution, we are looking for specific properties. Using the equations for pressure, volume, and temperature, we plot a $P-V$ diagrams to evaluate the critical behavior and the critical characteristics of the NLED AdS BH. From the Figure \textbf{16} and \textbf{19}, it is obvious that a thermodynamic system behaves like an ideal gas and is stable when $ T > T_{c} $. When $ T = T_{c} $, the critical isotherm has an inflection point, and when $ T <T_{c} $ the unstable zone exists in the system and we observe that the small/large BH phase transition occurs. To understand more clearly, a phase transition is displayed in right plots in Figs. $17$, and $20$. It may noted that $P-V$ criticality have stable and unstable regions. The stability can be seen in the regions $V\in[0,a]$ and $[b, \infty]$ corresponding to the small and large BHs, respectively. While there is un-stability in the region $V \in[a, b]$ and the phase transition can coexist for the unstable region of small and large BH. To show the impact of NLED parameter $\alpha$ in $P-V$ criticality is plotted in Figs.$18$, and $21$ to show phase transition. The $P-V$ diagram in Figs. \textbf{16} and \textbf{19} shows that the stationary points of inflection are at the same location where the critical points of equation of the state occur.
Now to determine the critical values, we differentiate Eq.(\ref{a8}) and get
\begin{equation}
r_{c}=\frac{\sqrt{6} Q}{\sqrt{2 \alpha  Q+1}},\label{a14}
\end{equation}
\begin{equation}
V_{c}=\frac{2 \left(\sqrt{6} Q\right)}{\sqrt{2 \alpha  Q+1}}, \label{a15}
\end{equation}
\begin{equation}
T_{c}= \frac{\frac{4 \sqrt{6} \alpha ^2 Q^2}{\sqrt{2 \alpha  Q+1}}+\frac{4 \sqrt{6} \alpha  Q}{\sqrt{2 \alpha  Q+1}}+\frac{\sqrt{6}}{\sqrt{2 \alpha  Q+1}}}{18 \pi  Q}, \label{a16}
\end{equation}
\begin{equation}
P_{c}=\frac{9 Q^3 \sqrt{2 \alpha  Q+1}+9 \alpha  Q r_{+}^4 \sqrt{2 \alpha  Q+1}+2 \sqrt{6} r_{+}^3 (2 \alpha  Q+1)^2-9 Q r_{+}^2 (2 \alpha  Q+1)^{3/2}}{72 \pi  Q r_{+}^4 \sqrt{2 \alpha  Q+1}}. \label{a17}
\end{equation}
Further, using Eqs.(\ref{a15}), (\ref{a16}) and (\ref{a17}), we have $ V_{c}=2.14834$, $T_{c}=\frac{0.4031}{\pi } $, and $P_{c}= 0.034350. $
The next step is to investigate the critical exponents, which demonstrate the feature of phase transitions generally. In the vicinity of critical point, there are four critical exponents that describe specific heat, parameter order, isothermal compressibility, and the critical isotherm, $\alpha$, $\beta$, $\gamma$ and $\delta$ which are frequently used to define phase transition as like the Van der Waals.
We define
$$ \tau= \frac{T}{T_{c}}-1 = t-1, $$
\\ where,
\begin{equation}
t=\frac{T}{T_{c}},
\end{equation}
\begin{equation}
\omega = \frac{v-v_{c}}{v} = \frac{v}{v_{c}}-1 = \phi -1,
\end{equation}
\\ where,\,\,\,\,\,\,\,\,
$$ \phi = \frac{v}{v_{c}}. $$
$$ p = \frac{P}{P_{c}}, \,\,\,\,\,\,\,\Rightarrow P=pP_{c}. $$
Identifying the critical exponents as $ t=1+ \tau \,\,\, \phi= 1+\omega. $
\begin{equation}
p\approx 0.9514+ {\tau}-{\tau \omega}- {1.972\omega^{3}}+ O(\tau \omega^{2}, \omega^{4} ).\label{a11}
\end{equation}
Black Hole sustain a step change from small to large, besides temperature, pressure, and thermodynamic volume remain constant.
The equation of state Eq. (\ref{a11})\,\,holds
\begin{equation}
p = 0.9514+ {\tau}-{\tau \omega_{s}}- {1.972\omega_{s}^{3}}= 0.9514+ {\tau}-{\tau \omega_{l}^{3}}- {1.972\omega_{l
}^{3}}.
\end{equation}
The Maxwell's area law ($ \oint v dp=0 $) during phase transition hold,
\begin{equation}
\int_{\omega_{s}}^{\omega_{l}} \omega dp= \int_{\omega_{s}}^{\omega_{l}}(\omega(p_{c}+5.916 \omega^{2}))\omega dp,
\end{equation}
After certain calculation, one can get
\begin{equation}
\omega_{l}=-\omega_{s},
\end{equation}
and we obtain,
$$ \omega_{l}=\sqrt{1.333 \tau}. $$
$1-$ Specific heat at constant value $S = \pi r^{2}$ is governed by $"\alpha_{1}"$
\begin{equation}
C_{v}= T \frac{\partial S}  {\partial T}|_{v}  \varpropto |\tau|,^{-\alpha_{1}}
\end{equation}
$$ \Rightarrow \frac{\partial S}{\partial T}=0. $$
As, entropy $S$ is autonomously from $T$ so, we finalize that $\alpha_{1} = 0$

$ 2- $ Exponent $\beta$ describes order parameter $\eta = V_{s} - V_{l}$.
\begin{equation}
\eta = V^{s}- V_{l}  \varpropto |\tau|,^{\beta}
\end{equation}
which simplifies to
$$   \eta = V_{s} - V_{l} = V_{c}(1+\omega_{s})- V_{c}(1+\omega_{l}),$$
$$ \eta = |\tau|^{\frac{1}{2}}, $$
thus
\begin{equation}
 \Rightarrow \beta = \frac{1}{2}.
\end{equation}
$ 3- $ Exponent $\gamma$ determine the isothermal compressibility $k\tau$ from Eq. ( \ref{a11}), we have
\begin{equation}
k\tau = \frac{-1}{V} \frac{\partial V}{\partial P}|_{T} \varpropto |\tau|^{-\gamma},
\end{equation}
\begin{equation}
k\tau = \frac{-1}{V} \frac{\partial V}{\partial P}|_{T} \varpropto |\tau|^{-1},
\end{equation}
thus
 \begin{equation}
\Rightarrow \gamma = 1.
\end{equation}
$ 4- $ As $T=T_{c}$, in a critical isotherm, which is given below
\begin{equation}
|P-P_{c}|_{T_{c}} \varpropto |V-V_{c}|^{\delta},
\end{equation}
\begin{equation}
\alpha_{1} + \beta (\delta +1) = 2,
\end{equation}
which implies that ~~~$\delta = 3$.
Hence, the critical elements connected to BH are helpful to investigate the critical behavior, phase transition, and critical exponents.
\section{Final remarks}
In this research, we have explored the phase structure and critical behavior of BH as well as the extended thermodynamics for AdS charged BH with NLED by using thermodynamic pressure as the cosmological constant. The mass$M$, charge $Q$, coupling constant $\alpha$  and cosmological constant define the AdS charged BH.

We investigate the features of such NLED BH in order to determine their stability and transition by computing the appropriate thermodynamical statistics.The horizon of the charged AdS BH solution is obtained by considering $N(r)=0$, which demonstrates that as the electric charge $Q$ improves, the BH size shrinks with a fixed value of $\alpha$ and $l$. When we study the Hawking temperature, we can see that as the electric charge decreases, the temperature rises to a maximum as the electric charge diminishes. In the absence of electric charge $Q$, it gives the Schwarzschild AdS BH temperature. Black hole entropy is calculated in accordance with the first law. The heat capacity with cosmological constant, which reveals information about the stability plotted in fig.$7$ and $8$ for different values of charge with fixed $\alpha$ and $l$. The heat capacity for the specified charge values is perfectly discontinuous at the critical values, as can be seen. Heat capacity determine stability for small radius BHs and show instability for large radius BHs. When a BH undergoes a phase change, its thermodynamic stability is described by the GFE. It is significant to investigate the fundamental characteristics and behavior of the BH. With the negative values of pressure at $p=-0.1,-0.3$ GFE express the stability of the thermodynamical system and remain stable for large radius BHs.
Pressure with positive values indicate the unstable system. In order to give information about critical aspects, we investigate the associated P-V depiction in fig.11, and 12. When $T<T_{c}$, one may have unstable region,when the critical isotherm indicates a point of inflection. Finally when $T>T_{c}$, the system behave like an ideal gas. Critical exponents, which describe how physical variables behave when they are close to a continuous phase transition. Critical exponent values are found to rise with charge and fall with the coupling constant.

\end{document}